\documentclass[preprints,article,accept,oneauthor,pdftex]{Definitions/mdpi}

\firstpage{1} 
\pubvolume{xx}
\issuenum{1}
\articlenumber{5}
\pubyear{2020}
\copyrightyear{2020}
\history{}

\pdfoutput=1

\Title{Explainable Ensemble Machine Learning for Breast Cancer Diagnosis based on Ultrasound Image Texture Features}

\Author{Alireza Rezazadeh $^{1}$, Yasamin Jafarian $^{2}$ and Ali Kord {\scriptsize
MD, MPH} $^{3}$}

\AuthorNames{Alireza Rezazadeh, Yasamin Jafarian and Ali Kord, MD, MPH}

\address{%
$^{1}$ \quad Department of Electrical and Computer Engineering, University of Minnesota; rezaz003@umn.edu\\
$^{2}$ \quad Department Computer Science and Engineering, University of Minnesota; yasamin@umn.edu\\
$^{3}$ \quad Division of Interventional Radiology, Department of Radiology, University of Cincinnati; kordai@uc.edu
}

\corres{Correspondence: rezaz003@umn.edu}

\abstract{Image classification is widely used to build predictive models for breast cancer diagnosis. Most existing approaches overwhelmingly rely on deep convolutional networks to build such diagnosis pipelines. These model architectures, although remarkable in performance, are black-box systems that provide minimal insight into the inner logic behind their predictions. This is a major drawback as the explainability of prediction is vital for applications such as cancer diagnosis. In this paper, we address this issue by proposing an explainable machine learning pipeline for breast cancer diagnosis based on ultrasound images. We extract first- and second-order texture features of the ultrasound images and use them to build a probabilistic ensemble of decision tree classifiers. Each decision tree learns to classify the input ultrasound image by learning a set of robust decision thresholds for texture features of the image. The decision path of the model predictions can then be interpreted by decomposing the learned decision trees. Our results show that our proposed framework achieves high predictive performance while being explainable. }

\begin{document}



\section{Introduction}
Ultrasound imaging is an effective method for breast cancer diagnosis \cite{kuhl2005mammography, al2020dataset} that, compared to alternative modalities, is more accessible and less costly. Several recent studies have explored building data-driven automated breast cancer diagnosis machine learning pipelines to detect the malignancy of tumors observed in ultrasound images \cite{moon2020computer, samulski2010using, sahiner2007malignant, jimenez2020deep}. These studies dominantly rely on deep convolutional neural network architectures to classify tumor images. Convolutional neural networks essentially learn to map the input image pixel information to a lower-dimensional feature space through a series of hidden layers. 

Although notable in prediction performance, convolutional networks are largely black-box machine learning models that provide little to no insight into the logic behind their predictions \cite{castelvecchi2016can, arrieta2020explainable}. In general, humans tend to be unwilling to rely on procedures that are not interpretable, explainable and transparent, especially for making critical predictions such as cancer diagnosis \cite{arrieta2020explainable, zhu2018explainable}. Yet, the explainability of machine learning models for cancer diagnosis falls short of the increasing demand for interpretable and reliable artificial intelligence \cite{preece2018stakeholders}. 

In a recent study, Moon et al. \cite{moon2020computer} adopted standard deep convolutional neural network architectures (including VGG, ResNet, and DenseNet) to classify breast ultrasound images and detect the malignancy of tumors. The study reports a high predictive performance for these standard architectures. In a similar study, Masud et al. \cite{masud2020convolutional} evaluated pretrained convolutional models for ultrasound image classification. Other studies, such as \cite{byra2020breast, irfan2021dilated}, focused on semantic segmentation of the breast tumor from ultrasound images.

In this paper, we propose an explainable machine learning pipeline for probabilistic breast cancer diagnosis based on ultrasound images. We formulate this as a binary classification problem. First, through a comprehensive texture analysis, we extract first- and second-order texture features of the region of interest in the ultrasound image. We then use these features to train an ensemble of decision trees. Each decision tree in the model learns to classify the input image through a set of test conjunctions where each test compares a texture feature with a robust numerical decision threshold. We show that our proposed pipeline achieves a high predictive performance that is comparable to the existing black box convolution neural network architectures. More importantly, we demonstrate that our proposed model can be probed to accurately track and explain the decision path behind its prediction.

\section{Materials and Methods}
\subsection{Data}
We use a public dataset of breast ultrasound images \cite{al2020dataset}. In this dataset, a total number of 780 images are obtained from 600 female patients (age of 25-75 years old). This includes 133 normal cases with no mass, 210 cases with a benign mass, and 487 cases with a malignant mass. Images are obtained using LOGIQ E9 ultrasound and LOGIQ E9 Agile ultrasound systems. These instruments produce DICOM images with $1280\times1024$ resolution using 1-5 MHz transducers on ML6-15-D Matrix linear probe. The raw DICOM images are cropped, preprocessed, and converted to PNG format with an average resolution of $500\times500$ pixels. For each case with a mass, a ground-truth binary mask of the region of interest (ROI) is manually created. Dataset is split into $80\%$ train set and $20\%$ test set.

\subsection{Texture Analysis}
Texture features are important quantifiable metrics to characterize and describe a region of interest in an image \cite{tuceryan1993texture, materka1998texture, varghese2019texture}. Texture feature analysis is typically measured using first-order and second-order statistical metrics \cite{srinivasan2008statistical}. 

\subsubsection{First-Order Statistics Texture Features}
First-order texture features are computed based on first-order statistics of the one-dimensional gray level histogram of the image. Therefore, a first-order texture feature does not take into account the pixel neighborhood information \cite{srinivasan2008statistical, kim1998ultrasound}. In this study, we compute eight common first-order statics of the ROI pixels: Mean, Variance, RMS, Energy, Entropy, Kurtosis, Skewness, and Uniformity. These statistical metrics described in Table \ref{tab:fos}. 

\subsubsection{Second-Order Statistics Texture Features}
The second-order statistical texture features are computed based on the gray-level co-occurrence matrix (GLCM). GLCM elements are an estimation of the probability of transition from one gray level to another along a certain pixel distance and direction \cite{sebastian2012gray, iqbal2017texture, xu2019classification}. We measure five common statistics based on the GLCM computed for the ROI: Contrast, Dissimilarity, Homogeneity, Energy, and Correlation. A description of these statistics are included in Table \ref{tab:glcm-math}.

\subsection{Decision Tree Models}
A decision tree (DT) is a predictive model that consists of a set of test conjunctions, where each test compares a feature of data with a numerical threshold \cite{sharma2016survey}. Decision tree classification models are learned by recursively partitioning the feature space to discover a set of robust decision rules \cite{myles2004introduction, safavian1991survey}. One major advantage of decision tree modeling is their interpretability. The set decision rules learned in a decision tree model can be directly used to explain the logic behind the prediction of the model. Explainable predictions, as opposed to black-box predictions, can be used more reliably in applications such as medical diagnosis. 

\subsubsection{Gradient Boosting Decision Tree}
Gradient boosting decision tree (GBDT) is an ensemble of sequential decision tree models. GBDT is frequently used in a variety of machine learning tasks due to its accuracy and efficiency. At each boosting iteration, the ensemble learns a decision tree to predict the residual errors \cite{sharma2016survey, ke2017lightgbm}. Specifically, for a classification problem, let $\{(X_i, y_i)\}_{i=1\dots N}$ be a dataset where for each entry, $X$ and $y$ correspond to feature vector and the class label the entry belongs to. The goal is to approximate the function $\hat{F}(X)=y$ that learns to map feature vectors to their corresponding class label under an arbitrary differentiable loss function $L(y_i, F(X_i))$, cross-entropy loss in our case. 

With gradient boosting, at first the model is initialized with a constant,
\begin{equation}
    F_0(x) = \min_{\lambda}\sum_{i=1}^{n}{L(y_i, \lambda)}
\end{equation}
Next, at each boosting iteration $m$, for each entry the residuals are computed as 
\begin{equation}
    r_{i, m} = - \partial L(y, F)/\partial F|_{F=F_{m-1}} 
\end{equation}
A decision tree with $J$ terminal nodes is fitted to the residuals where
\begin{equation}
    \lambda_{j,m} = \min_{\lambda}\sum_{j}{L(y, F_{m-1}(x)+\lambda)}
\end{equation}
Finally, the boosted model is updated as:
\begin{equation}
    F_m(x) = F_{m-1}(x)+\nu \sum_{J}{\lambda_{j,m}I(x)}
\end{equation}

Through grading boosting, GBDT combines multiple "weak" learner classifiers into an ensemble of strong classification model. 

\subsubsection{LightGBM}
LightGBM is an open-source GBDT framework \cite{ke2017lightgbm}. LightGBM is based on a gradient-based one-side sampling to filter data instances and an exclusive feature bundling to encode features into less dense space. Specifically, LightGBM discretizes continuous features using a histogram-based algorithm for a faster training process and reduced memory consumption. Also, LightGBM uses a leaf-wise strategy of growing decision trees by discovering a leaf with the highest gain of variance. This enables LightGBM to achieve state-of-the-art performance in a variety of applications \cite{rezazadeh2020generalized, chen2019lightgbm, sun2020novel}.

\begin{figure}
\centering
\includegraphics[width=13 cm]{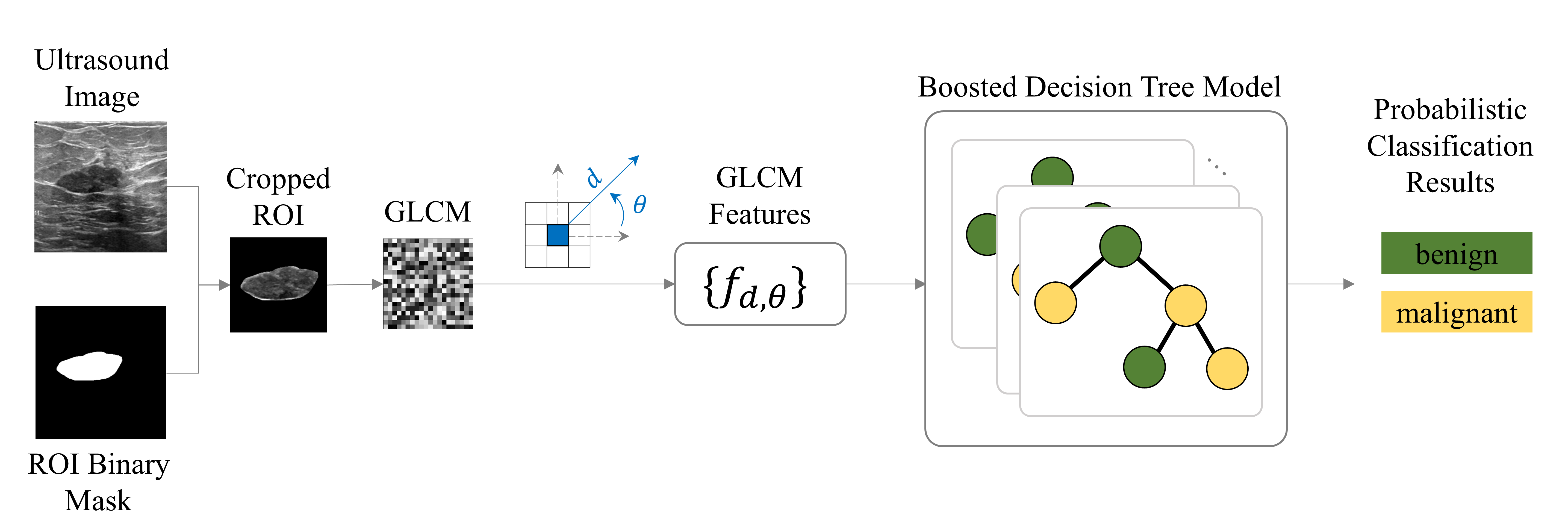}
\caption{Our proposed pipeline uses the ultrasound image and ROI mask to extract GLCM texture features and learns a boosted decision tree model to make probabilistic diagnosis based on explainable decision trees. \label{fig:model}}
\end{figure}   

\subsection{Machine Learning Diagnosis Pipeline}
In this paper, we propose an interpretable machine learning pipeline for breast cancer diagnosis based on ultrasound images. We formulate this as a binary classification problem where class labels belong to \{benign, malignant\}. As shown in Figure \ref{fig:model}, the pipeline takes as input the ultrasound image of the mass and a binary mask of the ROI. We use this binary mask to crop and compute the GLCM of the ROI. This yields a total number of 60 texture features ($5$ GLCM statistics $\times 3$ distances $\times 4$ angles). These texture features are denoted as \textit{\{statistic\}\_d\{distance\}\_a\{angle\}} (e.g., \textit{energy\_d3\_a135} refers to the GLCM energy computed within a distance of 3 pixels and along a 135$^{\circ}$ direction). 

We then use a LightGBM classification model with a gradient boosting decision tree strategy, 10 leaves per tree, and a maximum feature bin size of 512. The classification model is trained by minimizing a binary log loss, with a learning rate of 0.05, and for a total number of 500 boosting iterations. For a given ultrasound image input, the model outputs the probability of the mass in ROI being benign or malignant. Importantly, the decision tree ensemble can be decomposed to explain how the model comes up with a prediction. The learned model is a set of decision trees with multiple test conjunctions that compare the texture features of the ROI with numerical thresholds inferred from the data. 

\section{Results}
In this section, we first summarize the texture analysis results and then evaluate the performance of our purposed pipeline. Lastly, we highlight how our pipeline can be used as an explainable machine learning framework to understand the logic behind each of its diagnostic predictions.

\subsection{Texture Analysis Statistical Analysis}
We perform texture analysis by computing the ROI first- and second-order statistics (see Table \ref{tab:fos} and \ref{tab:glcm-math} for mathematical descriptions). A standard t-test is used to compare each texture feature between the two groups of benign and malignant masses. 

\begin{figure}
\centering
\includegraphics[width=12.5 cm]{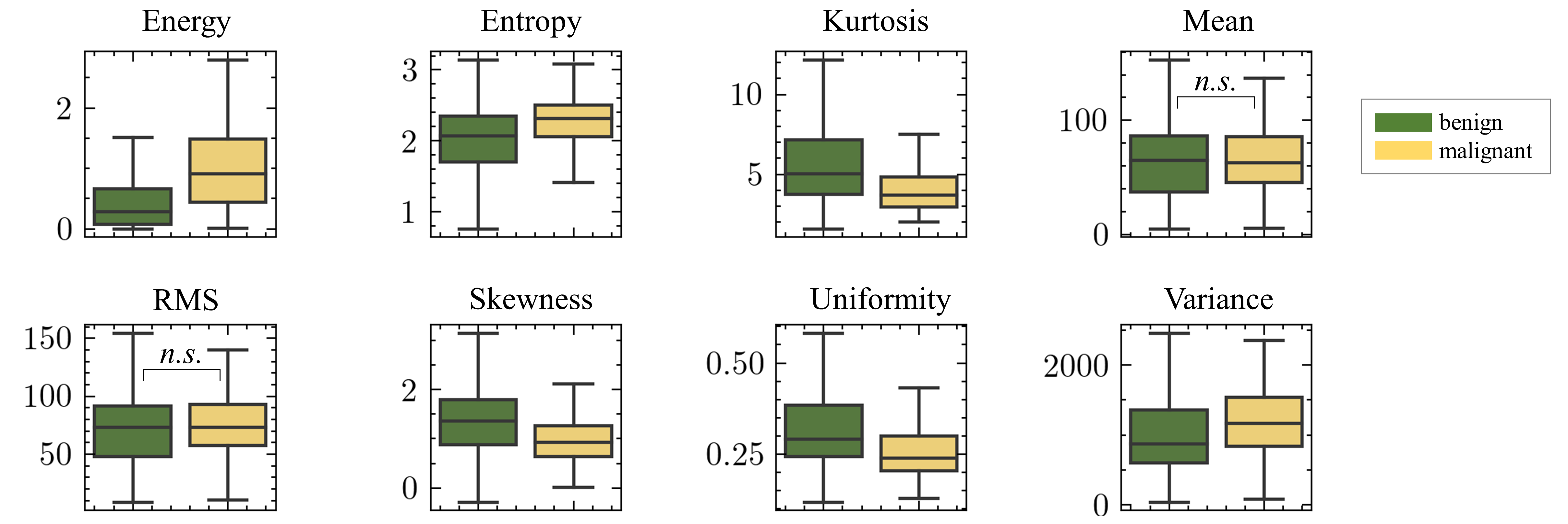}
\caption{First-order texture features. All measured statistics, except Mean and RMS, are significantly different between the two groups of benign and malignant masses. Refer to Table \ref{tab:fos} for a description of the metrics and t-test results. \label{fig:fos}}
\end{figure}

Most first-order statistic texture features are significantly different for the two sets of benign and malignant masses ($p<0.001$). Mean ($p=0.43$) and RMS ($p=0.28$) first-order statistics, however, are not significantly different across the groups. Figure \ref{fig:fos} demonstrates the first-order texture features comparison. All t-test results of this comparison are thoroughly reported in Table \ref{tab:fos}.

\begin{figure}
\centering
\includegraphics[width=12 cm]{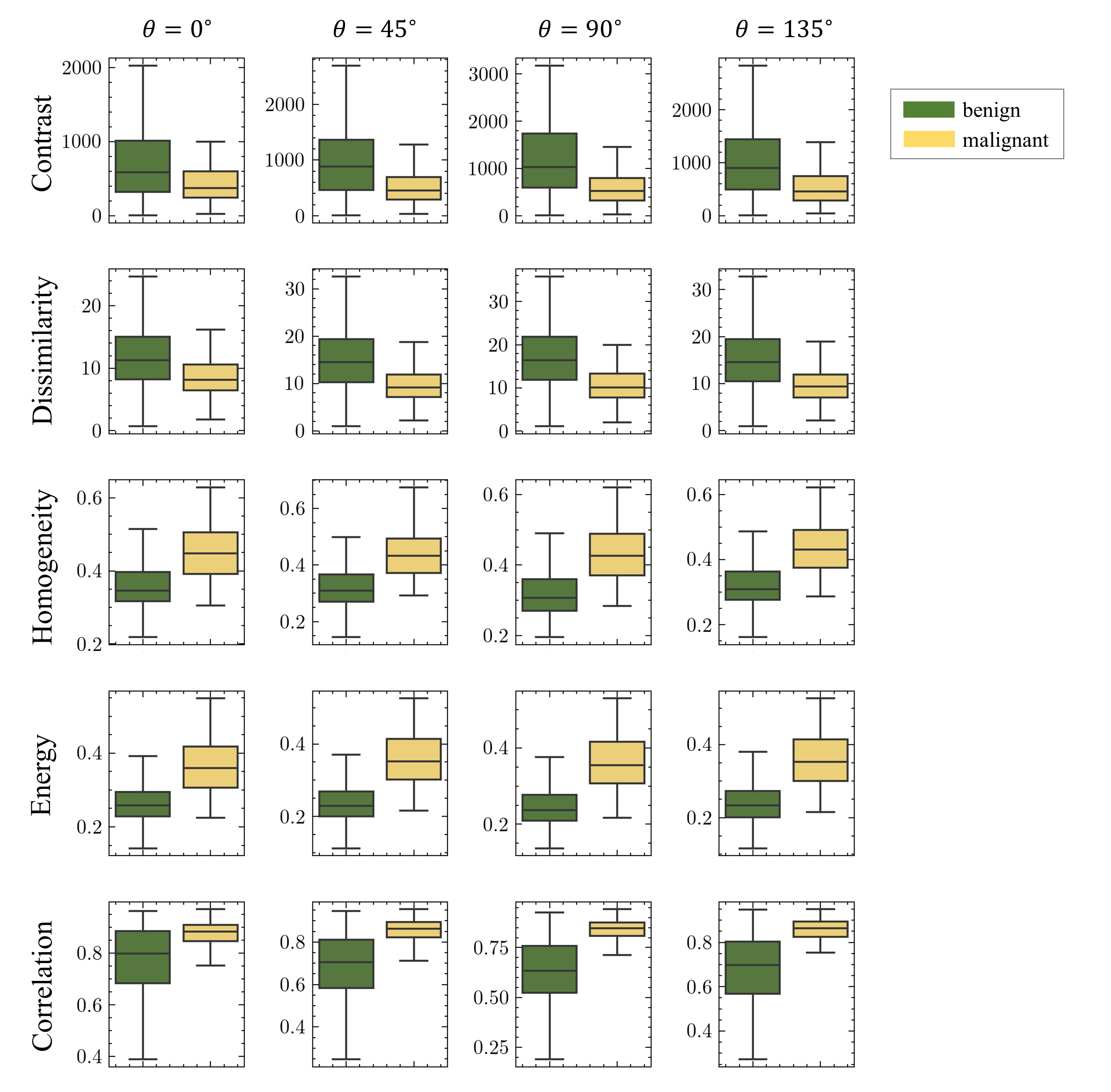}
\caption{Second-order GLCM texture features. All measure statistics are significantly different between the benign and malignant groups. This significant difference is consistent across various pixel distances ($d=\{1,3,5\}$, only d=3 results are demonstrated here) and angles ($\theta=\{0^{\circ}, 45^{\circ}, 90^{\circ}, 135^{\circ}\}$). Refer to Tables \ref{tab:glcm-math}-\ref{tab:glcm-homo} for the deception of each metric and detailed t-test results.\label{fig:glcm}}
\end{figure}

Second-order statics based on GLCM are computed across 3 distances $\{1,3,5\}$ pixels and 4 angles $\{0^{\circ}, 45^{\circ}, 90^{\circ}, 135^{\circ}\}$. All GLCM features are significantly different between benign and malignant groups ($p<0.001$). See Tables \ref{tab:glcm-math}-\ref{tab:glcm-homo} for the complete t-test results. Interestingly, the difference between the two groups is consistent for various distances and angles. This further indicates that GLCM features are consistent for different orientations of the ROI. Figure \ref{fig:glcm} exemplifies the persistence of the difference in each GLCM features across all angles for $d=3$ pixels.

\subsection{Machine Learning Diagnosis Pipeline Performance}
In this section, we compare multiple variations of our framework. The comparison is made using common standard statistical metrics for evaluating classification performance on the test set. Specifically, we report: Accuracy, Precision, Recall, F1-score, and Area under the ROC Curve (AUC) (see \cite{hossin2015review} for mathematical descriptions of these metrics). 

Using LightGBM algorithm with 500 iterations of gradient boosting, our pipeline reaches its best performance with 0.91 accuracy, 0.94 precision, 0.93 recall, and 0.93 F1-score (Table \ref{tab:result}). With smaller number of boosting iterations, the accuracy slightly drops down. Also, compared to a simple DT, using LightGBM significantly improved the performance of our pipeline. This emphasizes the importance of gradient boosting approaches in our pipeline.  

\begin{table}
\caption{Model evaluation. Standard classification performance metrics measured on the test set. The best performance is achieved with LightGBM model with 500 gradient boosting iterations. \label{tab:result}}

\newcolumntype{C}{>{\centering\arraybackslash}X}
\centering
\begin{tabularx}{12 cm}{l|ccccc}
\toprule
Model   & Precision   & Recall   & F1-score   & AUC   & Accuracy   \\
\midrule
DT            & 0.85   & 0.82   & 0.83   & 0.82   & 0.86   \\
LightGBM-10   & 0.90   & 0.80   & 0.83   & 0.79   & 0.87   \\
LightGBM-50   & 0.88   & 0.87   & 0.88   & 0.90   & 0.87   \\
LightGBM-100  & 0.93   & 0.91   & 0.92   & 0.92   & 0.90   \\
LightGBM-500  & \textbf{0.94}   & \textbf{0.93}   & \textbf{0.93}   & \textbf{0.93}   & \textbf{0.91}   \\

\bottomrule
\end{tabularx}
\end{table}

We also compare our model's performance with the standard convolutional neural network architectures from Moon et al. \cite{moon2020computer} which is also using a similar dataset to ours \cite{al2020dataset}. Our pipeline, based on decision tree ensembles, achieves comparable results to the convolutional network models (Table \ref{tab:result-compare}) while being explainable. Our model can be probed to trace the logic behind its predictions while the convolutional models do not provide any insight into the process behind their predictions. In comparison, our model achieves higher precision, recall, and F1-score. Note that F1-score is more suitable to assess models for their classification performance on imbalanced datasets.

\begin{table}
\caption{Model comparison with convolutional architectures. Standard classification performance metrics measured on the test set. Our explainable model based on decision trees achieves high predictive performance that is comparable to existing black box convolutional neural network architectures. \label{tab:result-compare}}
\newcolumntype{C}{>{\centering\arraybackslash}X}
\centering
\begin{tabularx}{14 cm}{l|ccccc}
\toprule
Model   & Precision   & Recall   & F1-score   & AUC   & Accuracy   \\
\midrule
VGG            & 0.75   & 0.76   & 0.76   & 0.87   & 0.85   \\
ResNet   & 0.89   & 0.89   & 0.89   & 0.96   & 0.91   \\
DenseNet   & 0.90   & 0.92   & 0.91   & \textbf{0.97}   & \textbf{0.94}   \\
Decision Tree Ensemble (ours)  & \textbf{0.94}   & \textbf{0.93}   & \textbf{0.93}   & 0.93   & 0.91\\
\bottomrule
\end{tabularx}
\end{table}

\begin{figure}
\centering
\includegraphics[width=10 cm]{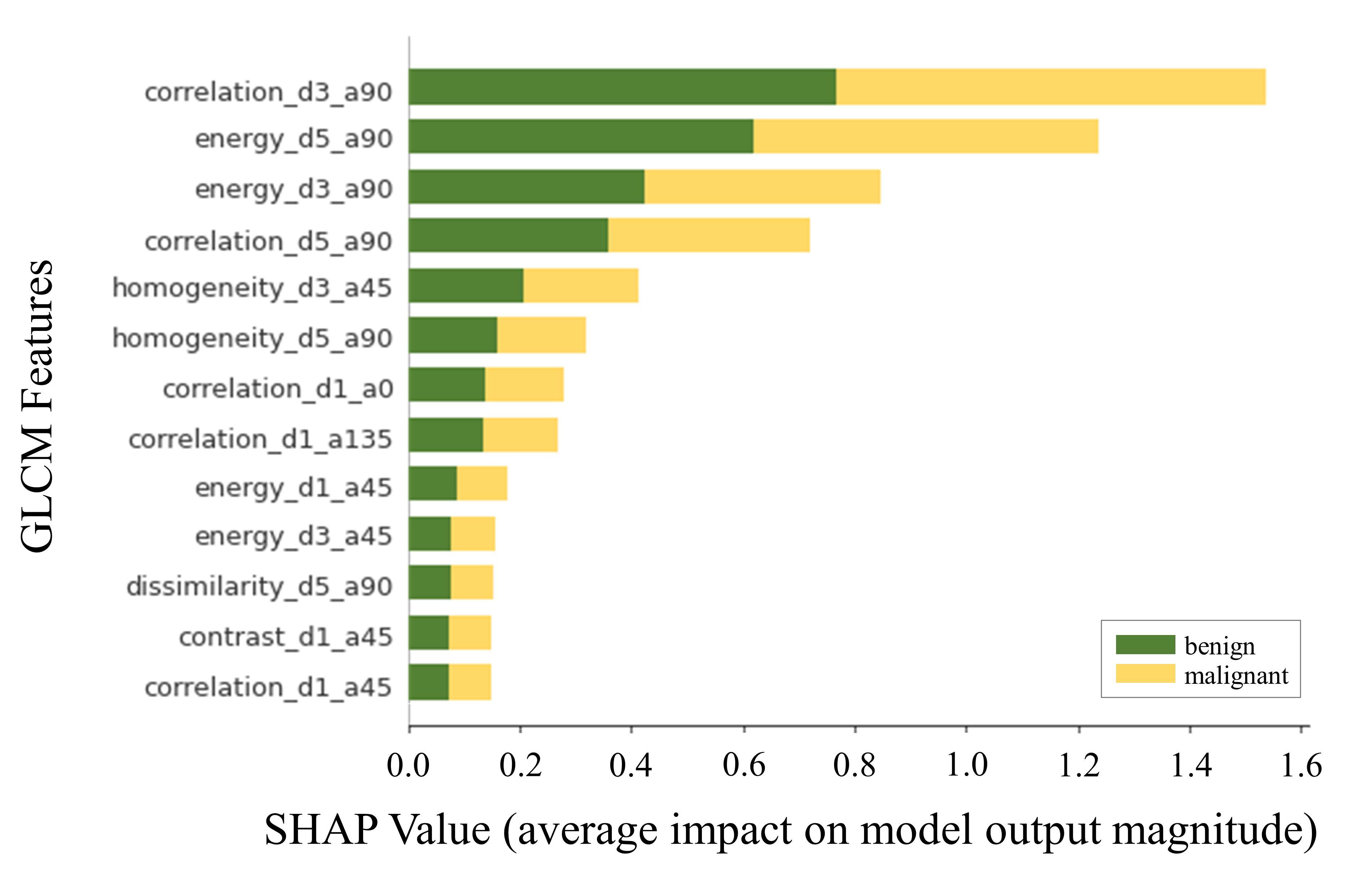}
\caption{Feature importance. SHAP value quantifies the contribution of each feature to the prediction of each class. The features are denoted as \textit{\{statistic\}\_d\{distance\}\_a\{angle\}}.\label{fig:featimport}}
\end{figure}

To further identify the most important texture features for the model, we quantify the feature importance with SHAP values \cite{lundberg2018consistent}. The SHAP values are Shapley values from coalitional game theory and correspond to the magnitude of each feature's attribution on the output of the model. With our dataset, the most important features are: GLCM correlation within 3 pixels along the 90$^{\circ}$ direction, GLCM energy within a 5 pixels distance along the 90$^{\circ}$ direction, GLCM energy within a 3 pixels distance along the 90$^{\circ}$ direction, and GLCM correlation within a 5 pixels distance along the 90$^{\circ}$ direction (see Figure \ref{fig:featimport}). Interestingly, all top 4 important features are statistics measured in the 90$^{\circ}$ direction. 

\subsection{Explainability of the Machine Learning Diagnosis Pipeline}
\begin{figure}
\centering
\includegraphics[width=\linewidth]{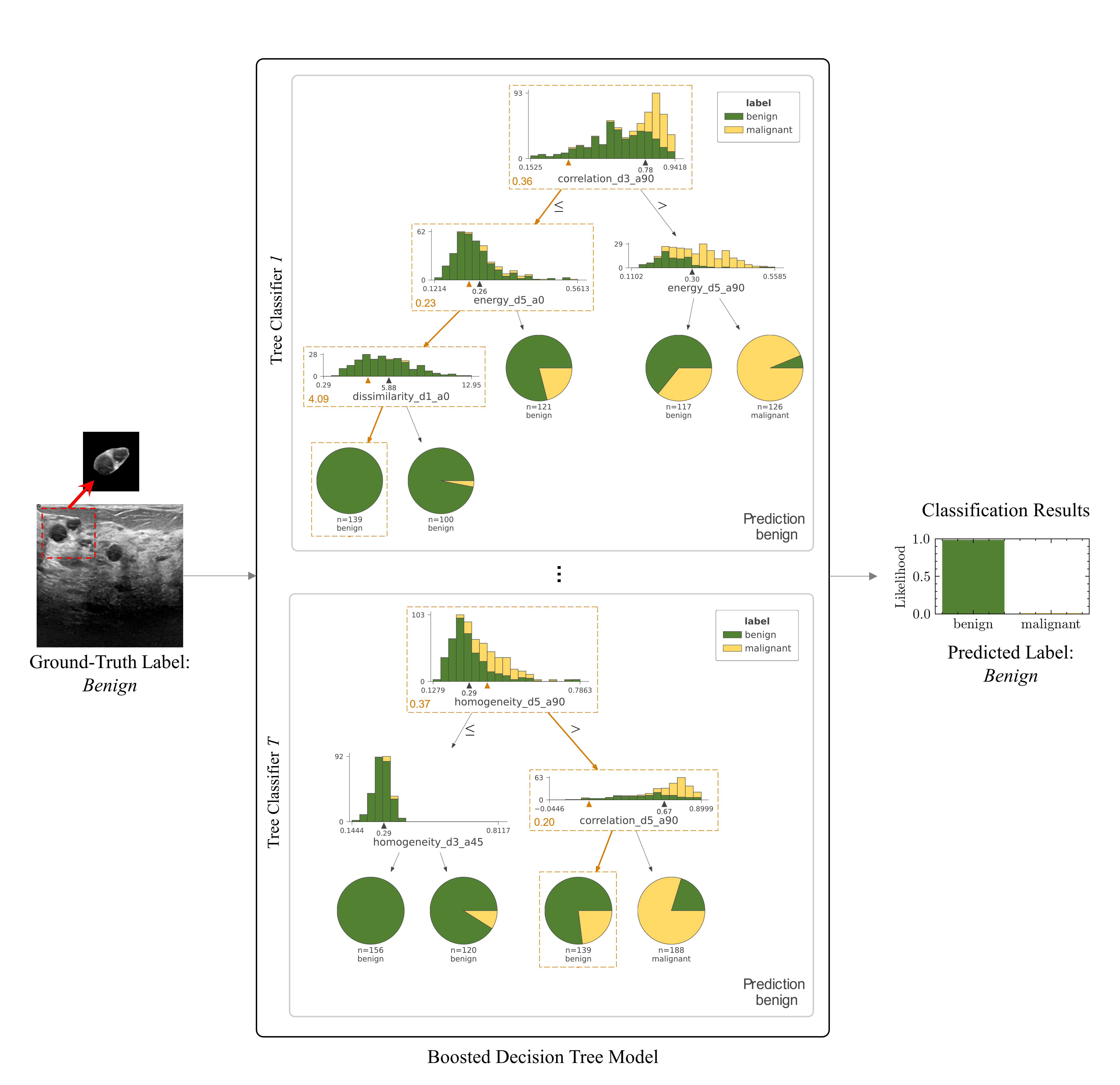}
\caption{Qualitative results of a benign case. Our pipeline learns to infer a probabilistic diagnosis of the breast ultrasound images. The learned ensemble can be probed to obtain explainable decision paths in a set of learned decision trees. In each learned tree classifier, the orange arrows highlight the decision path. The model learns to compare the texture features obtained from the input image (orange numbers at the bottom of each dashed box) with the learned thresholds (black triangle on each histogram) at each node of the decision tree. \label{fig:qual-ben}}
\end{figure}

In this section, we give an overview of how the explainability of our pipeline enables tracking down the logic behind its predictions. First, we evaluate a benign case from the test set (Figure \ref{fig:qual-ben}). Given the ultrasound image and the cropped ROI, the learned model predicts the mass in ROI is a benign with a likelihood of 0.97. We can further decompose the learned model to understand the decision path that leads to this prediction. The learned model is an ensemble of multiple decision trees ($T=500$).

In this example (Figure \ref{fig:qual-ben}), two decision trees from the ensemble are depicted. In \textit{Tree Classifier 1}, the first nodes splits the data based on $correlation\_d3\_a90$. This texture feature is measured as 0.36 in the cropped ROI which, in comparison with the learned split threshold 0.78, determines the outcome of the split at this node. The histogram of nodes also compares the split threshold with the distribution of the texture feature for both malignant and benign cases in the train set. At the next node $energy\_d5\_a0=0.23<0.26$ which leads to the final node $dissimilarity\_d1\_a0=4.09<5.88$. Based on this decision path the model predicts the ROI belongs to the benign class. Other decision trees in the ensemble can also be interpreted in a similar way.

\begin{figure}
\centering
\includegraphics[width=\linewidth]{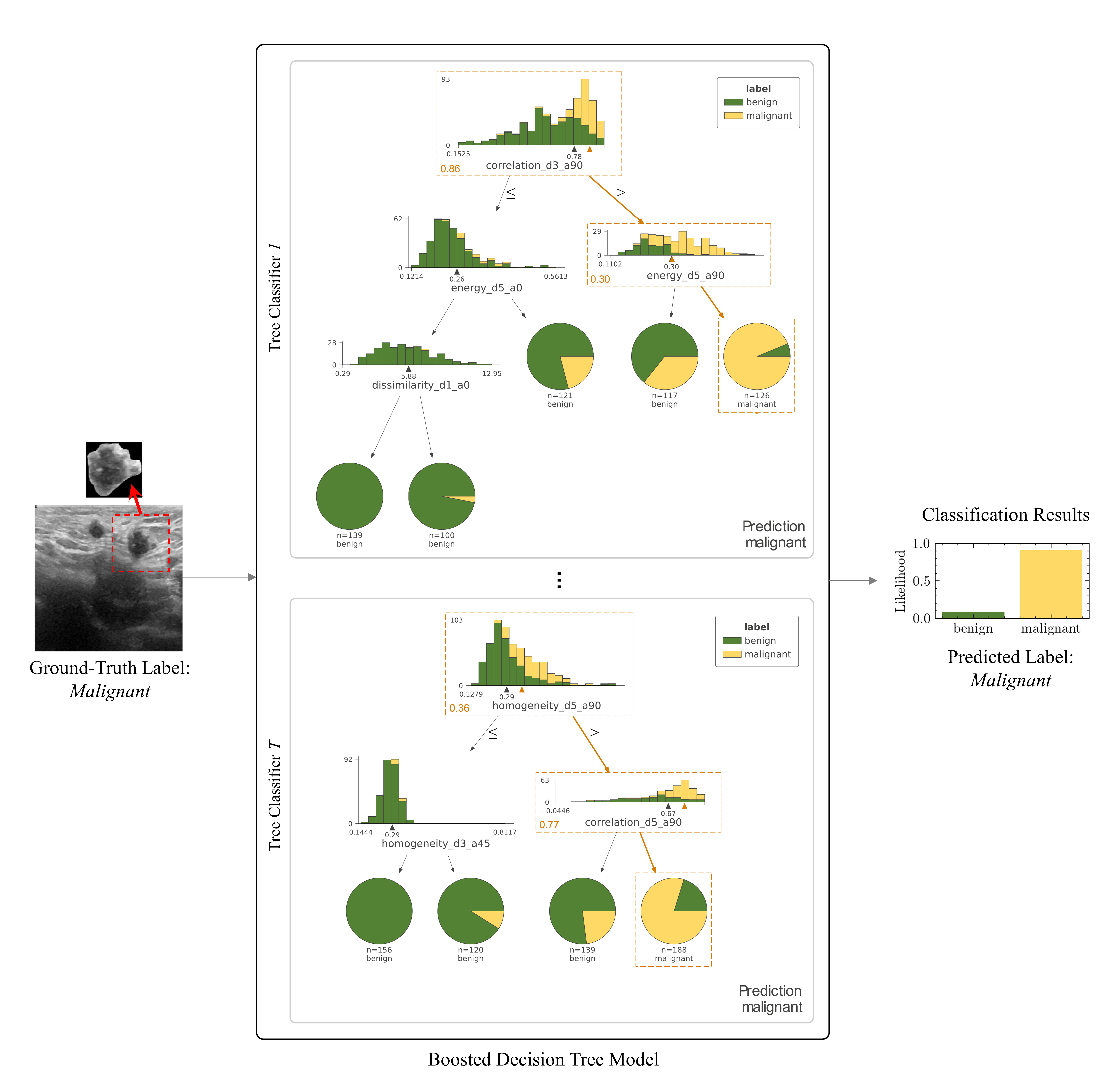}
\caption{Qualitative results of a malignant case. Refer to Figure \ref{fig:qual-ben} caption for more details. \label{fig:qual-malig}}
\end{figure}

Next, we evaluate a malignant case from the test set (Figure \ref{fig:qual-malig}). For this input, the model predicts the mass in ROI is malignant with a likelihood of 0.93. In \textit{Tree Classifier 1}, the model's decision path begins with a split based on $correlation\_d3\_a90=0.86>0.78$. This is followed by the final node $energy\_d5\_a90=0.31>0.30$ which predicts the ROI mass is malignant. 

\section{Conclusions}

Ultrasound imaging is an accessible and cost-effective imaging modality to diagnose breast cancer. Most of recent work on building machine learning models for breast cancer diagnosis depend on convolutional neural network architectures. Although accurate in performance, convolutional networks are black-box models and cannot be interpreted in terms of the logic behind their predictions.

In this paper, we proposed a novel explainable machine learning pipeline for breast cancer diagnosis based on ultrasound images. Our pipeline uses texture analysis of the ultrasound images as its basis to learn an ensemble of decision trees to predict the likelihood of malignancy of breast tumors. Importantly, our model can be decomposed into its underlying decision trees to fully interpret the decision path behind its outputs by following test conjunctions in each decision tree. 

We believe our work is a step towards building more practical and comprehensible machine learning tools for cancer diagnosis by increasing the transparency of the prediction process. An interesting future work is to combine convolutional networks with decision trees. Finally, we hope our approach in this work to inspire future research on data-driven medical diagnosis to devote more attention into increasing the explainability of their solutions.



\vspace{6pt} 

\appendixtitles{no} 
\appendix
\section[\appendixname~\thesection]{}
This section gives an overview of the first- and second-order GLCM texture feature statistics along with a detailed report of the t-test of each statistic between the two group of benign and malignant masses.

\begin{table}[H] 
\caption{First-order Texture Features. The metrics are computed for a set of $N$ pixels inside of the ROI denoted as $X$. Benign and malignant groups are compared using a t-test.\label{tab:fos}}
\centering
\newcolumntype{C}{>{\centering\arraybackslash}X}
\begin{tabular}{l|c|rrr}
\toprule
Feature & Description & $\mu_{benign}$ & $\mu_{malignant}$ & \textit{p-value} \\ \midrule
energy & $\sum^{N}_{i=1}{(\textbf{X}(i) + c)^2}$     & 6.66e+07 & 1.17e+08 & \textless 0.001 \\
entropy & $-\displaystyle\sum^{N}_{i=1}{p(i)\log_2\big(p(i)+\epsilon\big)}$    & 1.98 & 2.23 & \textless 0.001 \\
kurtosis & $\frac{\frac{1}{N}\sum^{N}_{i=1}{(\textbf{X}(i)-\bar{X})^4}}
      {\left(\frac{1}{N}\sum^{N}_{i=1}{(\textbf{X}(i)-\bar{X}})^2\right)^2}$  & 6.50 & 4.27 & \textless 0.001 \\
mean & $\frac{1}{N}\displaystyle\sum^{N}_{i=1}{\textbf{X}(i)}$       & 6.40e+01 & 6.59e+01 & 0.43            \\
rms & $\sqrt{\frac{1}{N}\sum^{N}_{i=1}{(\textbf{X}(i) + c)^2}}$       & 7.26e+01 & 7.51e+01 & 0.28            \\
skewness & $\frac{\frac{1}{N}\sum^{N}_{i=1}{(\textbf{X}(i)-\bar{X})^3}}
      {\left(\sqrt{\frac{1}{N}\sum^{N}_{i=1}{(\textbf{X}(i)-\bar{X})^2}}\right)^3}$   & 1.45 & 9.92e-01 & \textless 0.001 \\
uniformity & $\displaystyle\sum^{N}_{i=1}{p(i)^2}$ & 3.40e-01 & 2.69e-01 & \textless 0.001 \\
variance & $\frac{1}{N}\displaystyle\sum^{N}_{i=1}{(\textbf{X}(i)-\bar{X})^2}$   & 1.04e+03 & 1.22e+03 & \textless 0.001 \\
\bottomrule
\end{tabular}
\end{table}
\unskip

\begin{table}[H] 
\caption{Second-order Texture Features. The metrics are computed based on the gray-level co-occurrence matrix for a set of $N$ gray level of pixels inside of the ROI.\label{tab:glcm-math}}
\centering
\begin{tabular}{l|c}
\toprule
GLCM Feature & Description \\ \midrule
contrast & $\displaystyle\sum^{N}_{i=1}\displaystyle\sum^{N}_{j=1}{(i-j)^2p(i,j)}$\label{tab:glcm-cont}\\
energy & $\sum^{N}_{i=1}{(\textbf{X}(i) + c)^2}$ \\
correlation & $\frac{\sum^{N}_{i=1}\sum^{N}_{j=1}{p(i,j)ij-\mu_x\mu_y}}{\sigma_x(i)\sigma_y(j)}$\\
dissimilarity & $\displaystyle\sum^{N}_{i=1}\displaystyle\sum^{N}_{j=1}{|i-j|p(i,j)}$\\
energy & $\displaystyle\sum^{N}_{i=1}\displaystyle\sum^{N}_{j=1}{\big(p(i,j)\big)^2}$\\
energy & $\displaystyle\sum^{N}_{i=1}\displaystyle\sum^{N}_{j=1}{\big(p(i,j)\big)^2}$\\
homogeneity & $\displaystyle\sum^{N}_{i=1}\displaystyle\sum^{N}_{j=1}{\frac{p(i,j)}{1+|i-j|}}$ \\

\bottomrule
\end{tabular}
\end{table}
\unskip

\begin{table}[H] 
\caption{Gray level co-occurrence matrix contrast feature measured across three pixel distances $\{1, 3, 5\}$ and four angles $\{0^{\circ}, 45^{\circ}, 90^{\circ}, 135^{\circ}\}$.}
\centering
\newcolumntype{C}{>{\centering\arraybackslash}X}
\begin{tabular}{crccc}
\hline
\multicolumn{5}{c}{GLCM contrast}                       \\ \hline
distance & angle   & $\mu_{benign}$            & $\mu_{malignant}$            & \textit{p-value}          \\ \hline
1 & 0\textdegree   & 184.86  & 114.20  & \textless 0.001 \\
  & 45\textdegree  & 433.20  & 238.06  & \textless 0.001 \\
  & 90\textdegree  & 304.20  & 154.26  & \textless 0.001 \\
  & 135\textdegree & 438.01  & 238.07  & \textless 0.001 \\ \hline
3 & 0\textdegree   & 730.45  & 493.19  & \textless 0.001 \\
  & 45\textdegree  & 1032.13 & 582.21  & \textless 0.001 \\
  & 90\textdegree  & 1229.20 & 660.83  & \textless 0.001 \\
  & 135\textdegree & 1049.76 & 582.42  & \textless 0.001 \\ \hline
5 & 0\textdegree   & 1136.95 & 830.26  & \textless 0.001 \\
  & 45\textdegree  & 1768.76 & 1132.49 & \textless 0.001 \\
  & 90\textdegree  & 1681.63 & 1040.57 & \textless 0.001 \\
  & 135\textdegree & 1801.47 & 1130.44 & \textless 0.001 \\ \hline
\end{tabular}
\end{table}
\unskip

\begin{table}[H] 
\caption{Gray level co-occurrence matrix correlation feature measured across three pixel distances $\{1, 3, 5\}$ and four angles $\{0^{\circ}, 45^{\circ}, 90^{\circ}, 135^{\circ}\}$.\label{tab:glcm-corr}}

\centering
\newcolumntype{C}{>{\centering\arraybackslash}X}
\begin{tabular}{crccc}
\hline
\multicolumn{5}{c}{GLCM correlation}                       \\ \hline
distance & angle   & $\mu_{benign}$            & $\mu_{malignant}$            & \textit{p-value}          \\ \hline
1 &       0\textdegree  &  0.94 &  0.97 &  \textless 0.001 \\
  &      45\textdegree  &  0.87 &  0.94 &  \textless 0.001 \\
  &      90\textdegree  &  0.91 &  0.96 &  \textless 0.001 \\
  &     135\textdegree  &  0.86 &  0.94 &  \textless 0.001 \\\hline
3 &       0\textdegree  &  0.77 &  0.87 &  \textless 0.001 \\
  &      45\textdegree  &  0.68 &  0.85 &  \textless 0.001 \\
  &      90\textdegree  &  0.63 &  0.83 &  \textless 0.001 \\
  &     135\textdegree  &  0.68 &  0.85 &  \textless 0.001 \\\hline
5 &       0\textdegree  &  0.64 &  0.78 &  \textless 0.001 \\
  &      45\textdegree  &  0.46 &  0.70 &  \textless 0.001 \\
  &      90\textdegree  &  0.47 &  0.73 &  \textless 0.001 \\
  &     135\textdegree  &  0.44 &  0.70 &  \textless 0.001 \\\hline
\end{tabular}
\end{table}
\unskip

\begin{table}[H] 
\caption{Gray level co-occurrence matrix dissimilarity feature measured across three pixel distances $\{1, 3, 5\}$ and four angles $\{0^{\circ}, 45^{\circ}, 90^{\circ}, 135^{\circ}\}$.\label{tab:glcm-dissim}}
\centering
\newcolumntype{C}{>{\centering\arraybackslash}X}
\begin{tabular}{crccc}
\hline
\multicolumn{5}{c}{GLCM dissimilarity}                       \\ \hline
distance & angle   & $\mu_{benign}$            & $\mu_{malignant}$            & \textit{p-value}          \\ \hline
1 &       0\textdegree  &  4.84 &  3.41 &  \textless 0.001 \\
  &      45\textdegree  &  8.67 &  5.75 &  \textless 0.001 \\
  &      90\textdegree  &  7.09 &  4.58 &  \textless 0.001 \\
  &     135\textdegree  &  8.69 &  5.74 &  \textless 0.001 \\\hline
3 &       0\textdegree  &  12.10 &  8.84 &  \textless 0.001 \\
  &      45\textdegree  &  15.33 &  10.07 &  \textless 0.001 \\
  &      90\textdegree  &  17.04 &  10.97 &  \textless 0.001 \\
  &     135\textdegree  &  15.40 &  10.06 &  \textless 0.001 \\\hline
5 &       0\textdegree  &  16.76 &  12.76 &  \textless 0.001 \\
  &      45\textdegree  &  23.10 &  15.95 &  \textless 0.001 \\
  &      90\textdegree  &  22.06 &  14.98 &  \textless 0.001 \\
  &     135\textdegree  &  23.18 &  15.90 &  \textless 0.001 \\\hline
\end{tabular}
\end{table}
\unskip

\begin{table}[H] 
\caption{Gray level co-occurrence matrix energy feature measured across three pixel distances $\{1, 3, 5\}$ and four angles $\{0^{\circ}, 45^{\circ}, 90^{\circ}, 135^{\circ}\}$.\label{tab:glcm-enrgy}}
\centering
\newcolumntype{C}{>{\centering\arraybackslash}X}
\begin{tabular}{crccc}
\hline
\multicolumn{5}{c}{GLCM energy}                       \\ \hline
distance & angle   & $\mu_{benign}$            & $\mu_{malignant}$            & \textit{p-value}          \\ \hline
1 &       0\textdegree  &  0.30 &  0.38 &  \textless 0.001 \\
  &      45\textdegree  &  0.28 &  0.37 &  \textless 0.001 \\
  &      90\textdegree  &  0.29 &  0.38 &  \textless 0.001 \\
  &     135\textdegree  &  0.28 &  0.37 &  \textless 0.001 \\\hline
3 &       0\textdegree  &  0.26 &  0.36 &  \textless 0.001 \\
  &      45\textdegree  &  0.24 &  0.35 &  \textless 0.001 \\
  &      90\textdegree  &  0.25 &  0.35 &  \textless 0.001 \\
  &     135\textdegree  &  0.24 &  0.35 &  \textless 0.001 \\\hline
5 &       0\textdegree  &  0.24 &  0.34 &  \textless 0.001 \\
  &      45\textdegree  &  0.19 &  0.32 &  \textless 0.001 \\
  &      90\textdegree  &  0.22 &  0.33 &  \textless 0.001 \\
  &     135\textdegree  &  0.19 &  0.32 &  \textless 0.001 \\\hline
\end{tabular}
\end{table}
\unskip

\begin{table}[H] 
\caption{Gray level co-occurrence matrix homogeneity feature measured across three pixel distances $\{1, 3, 5\}$ and four angles $\{0^{\circ}, 45^{\circ}, 90^{\circ}, 135^{\circ}\}$.\label{tab:glcm-homo}}
\centering
\newcolumntype{C}{>{\centering\arraybackslash}X}
\begin{tabular}{crccc}
\hline
\multicolumn{5}{c}{GLCM homogeneity}                       \\ \hline
distance & angle   & $\mu_{benign}$            & $\mu_{malignant}$            & \textit{p-value}          \\ \hline
1 &       0\textdegree  &  0.49 &  0.56 &  \textless 0.001 \\
  &      45\textdegree  &  0.41 &  0.49 &  \textless 0.001 \\
  &      90\textdegree  &  0.43 &  0.51 &  \textless 0.001 \\
  &     135\textdegree  &  0.41 &  0.50 &  \textless 0.001 \\\hline
3 &       0\textdegree  &  0.37 &  0.45 &  \textless 0.001 \\
  &      45\textdegree  &  0.33 &  0.44 &  \textless 0.001 \\
  &      90\textdegree  &  0.33 &  0.43 &  \textless 0.001 \\
  &     135\textdegree  &  0.33 &  0.44 &  \textless 0.001 \\\hline
5 &       0\textdegree  &  0.33 &  0.41 &  \textless 0.001 \\
  &      45\textdegree  &  0.26 &  0.32 &  \textless 0.001 \\
  &      90\textdegree  &  0.30 &  0.40 &  \textless 0.001 \\
  &     135\textdegree  &  0.27 &  0.39 &  \textless 0.001 \\\hline
\end{tabular}
\end{table}
\unskip


\reftitle{References}


\externalbibliography{yes}
\bibliography{ref.bib}

\begin{thebibliography}{-------}
\providecommand{\natexlab}[1]{#1}

\bibitem[Kuhl \em{et~al.}(2005)Kuhl, Schrading, Leutner, Morakkabati-Spitz,
  Wardelmann, Fimmers, Kuhn, and Schild]{kuhl2005mammography}
Kuhl, C.K.; Schrading, S.; Leutner, C.C.; Morakkabati-Spitz, N.; Wardelmann,
  E.; Fimmers, R.; Kuhn, W.; Schild, H.H.
\newblock Mammography, breast ultrasound, and magnetic resonance imaging for
  surveillance of women at high familial risk for breast cancer.
\newblock {\em Journal of clinical oncology} {\bf 2005}, {\em 23},~8469--8476.

\bibitem[Al-Dhabyani \em{et~al.}(2020)Al-Dhabyani, Gomaa, Khaled, and
  Fahmy]{al2020dataset}
Al-Dhabyani, W.; Gomaa, M.; Khaled, H.; Fahmy, A.
\newblock Dataset of breast ultrasound images.
\newblock {\em Data in brief} {\bf 2020}, {\em 28},~104863.

\bibitem[Moon \em{et~al.}(2020)Moon, Lee, Ke, Lee, Huang, and
  Chang]{moon2020computer}
Moon, W.K.; Lee, Y.W.; Ke, H.H.; Lee, S.H.; Huang, C.S.; Chang, R.F.
\newblock Computer-aided diagnosis of breast ultrasound images using ensemble
  learning from convolutional neural networks.
\newblock {\em Computer methods and programs in biomedicine} {\bf 2020}, {\em
  190},~105361.

\bibitem[Samulski \em{et~al.}(2010)Samulski, Hupse, Boetes, Mus, den Heeten,
  and Karssemeijer]{samulski2010using}
Samulski, M.; Hupse, R.; Boetes, C.; Mus, R.D.; den Heeten, G.J.; Karssemeijer,
  N.
\newblock Using computer-aided detection in mammography as a decision support.
\newblock {\em European radiology} {\bf 2010}, {\em 20},~2323--2330.

\bibitem[Sahiner \em{et~al.}(2007)Sahiner, Chan, Roubidoux, Hadjiiski, Helvie,
  Paramagul, Bailey, Nees, and Blane]{sahiner2007malignant}
Sahiner, B.; Chan, H.P.; Roubidoux, M.A.; Hadjiiski, L.M.; Helvie, M.A.;
  Paramagul, C.; Bailey, J.; Nees, A.V.; Blane, C.
\newblock Malignant and benign breast masses on 3D US volumetric images: effect
  of computer-aided diagnosis on radiologist accuracy.
\newblock {\em Radiology} {\bf 2007}, {\em 242},~716--724.

\bibitem[Jim{\'e}nez-Gaona \em{et~al.}(2020)Jim{\'e}nez-Gaona,
  Rodr{\'\i}guez-{\'A}lvarez, and Lakshminarayanan]{jimenez2020deep}
Jim{\'e}nez-Gaona, Y.; Rodr{\'\i}guez-{\'A}lvarez, M.J.; Lakshminarayanan, V.
\newblock Deep-Learning-Based Computer-Aided Systems for Breast Cancer Imaging:
  A Critical Review.
\newblock {\em Applied Sciences} {\bf 2020}, {\em 10},~8298.

\bibitem[Castelvecchi(2016)]{castelvecchi2016can}
Castelvecchi, D.
\newblock Can we open the black box of AI?
\newblock {\em Nature News} {\bf 2016}, {\em 538},~20.

\bibitem[Arrieta \em{et~al.}(2020)Arrieta, D{\'\i}az-Rodr{\'\i}guez, Del~Ser,
  Bennetot, Tabik, Barbado, Garc{\'\i}a, Gil-L{\'o}pez, Molina, Benjamins,
  et~al.]{arrieta2020explainable}
Arrieta, A.B.; D{\'\i}az-Rodr{\'\i}guez, N.; Del~Ser, J.; Bennetot, A.; Tabik,
  S.; Barbado, A.; Garc{\'\i}a, S.; Gil-L{\'o}pez, S.; Molina, D.; Benjamins,
  R.; others.
\newblock Explainable Artificial Intelligence (XAI): Concepts, taxonomies,
  opportunities and challenges toward responsible AI.
\newblock {\em Information Fusion} {\bf 2020}, {\em 58},~82--115.

\bibitem[Zhu \em{et~al.}(2018)Zhu, Liapis, Risi, Bidarra, and
  Youngblood]{zhu2018explainable}
Zhu, J.; Liapis, A.; Risi, S.; Bidarra, R.; Youngblood, G.M.
\newblock Explainable AI for designers: A human-centered perspective on
  mixed-initiative co-creation.
\newblock  2018 IEEE Conference on Computational Intelligence and Games (CIG).
  IEEE,  2018, pp. 1--8.

\bibitem[Preece \em{et~al.}(2018)Preece, Harborne, Braines, Tomsett, and
  Chakraborty]{preece2018stakeholders}
Preece, A.; Harborne, D.; Braines, D.; Tomsett, R.; Chakraborty, S.
\newblock Stakeholders in explainable AI.
\newblock {\em arXiv preprint arXiv:1810.00184} {\bf 2018}.

\bibitem[Masud \em{et~al.}(2020)Masud, Rashed, and
  Hossain]{masud2020convolutional}
Masud, M.; Rashed, A.E.E.; Hossain, M.S.
\newblock Convolutional neural network-based models for diagnosis of breast
  cancer.
\newblock {\em Neural Computing and Applications} {\bf 2020}, pp. 1--12.

\bibitem[Byra \em{et~al.}(2020)Byra, Jarosik, Szubert, Galperin,
  Ojeda-Fournier, Olson, O’Boyle, Comstock, and Andre]{byra2020breast}
Byra, M.; Jarosik, P.; Szubert, A.; Galperin, M.; Ojeda-Fournier, H.; Olson,
  L.; O’Boyle, M.; Comstock, C.; Andre, M.
\newblock Breast mass segmentation in ultrasound with selective kernel U-Net
  convolutional neural network.
\newblock {\em Biomedical Signal Processing and Control} {\bf 2020}, {\em
  61},~102027.

\bibitem[Irfan \em{et~al.}(2021)Irfan, Almazroi, Rauf,
  Dama{\v{s}}evi{\v{c}}ius, Nasr, and Abdelgawad]{irfan2021dilated}
Irfan, R.; Almazroi, A.A.; Rauf, H.T.; Dama{\v{s}}evi{\v{c}}ius, R.; Nasr,
  E.A.; Abdelgawad, A.E.
\newblock Dilated semantic segmentation for breast ultrasonic lesion detection
  using parallel feature fusion.
\newblock {\em Diagnostics} {\bf 2021}, {\em 11},~1212.

\bibitem[Tuceryan and Jain(1993)]{tuceryan1993texture}
Tuceryan, M.; Jain, A.K.
\newblock Texture analysis.
\newblock {\em Handbook of pattern recognition and computer vision} {\bf 1993},
  pp. 235--276.

\bibitem[Materka \em{et~al.}(1998)Materka, Strzelecki,
  et~al.]{materka1998texture}
Materka, A.; Strzelecki, M.; others.
\newblock Texture analysis methods--a review.
\newblock {\em Technical university of lodz, institute of electronics, COST B11
  report, Brussels} {\bf 1998}, {\em 10},~4968.

\bibitem[Varghese \em{et~al.}(2019)Varghese, Cen, Hwang, and
  Duddalwar]{varghese2019texture}
Varghese, B.A.; Cen, S.Y.; Hwang, D.H.; Duddalwar, V.A.
\newblock Texture analysis of imaging: what radiologists need to know.
\newblock {\em American Journal of Roentgenology} {\bf 2019}, {\em
  212},~520--528.

\bibitem[Srinivasan and Shobha(2008)]{srinivasan2008statistical}
Srinivasan, G.; Shobha, G.
\newblock Statistical texture analysis.
\newblock  Proceedings of world academy of science, engineering and technology,
   2008, Vol.~36, pp. 1264--1269.

\bibitem[Kim \em{et~al.}(1998)Kim, Amin, Wilson, Rouse, and
  Udpa]{kim1998ultrasound}
Kim, N.D.; Amin, V.; Wilson, D.; Rouse, G.; Udpa, S.
\newblock Ultrasound image texture analysis for characterizing intramuscular
  fat content of live beef cattle.
\newblock {\em Ultrasonic imaging} {\bf 1998}, {\em 20},~191--205.

\bibitem[Sebastian~V \em{et~al.}(2012)Sebastian~V, Unnikrishnan, and
  Balakrishnan]{sebastian2012gray}
Sebastian~V, B.; Unnikrishnan, A.; Balakrishnan, K.
\newblock Gray level co-occurrence matrices: generalisation and some new
  features.
\newblock {\em arXiv preprint arXiv:1205.4831} {\bf 2012}.

\bibitem[Iqbal \em{et~al.}(2017)Iqbal, Pallewatte, and
  Wansapura]{iqbal2017texture}
Iqbal, F.; Pallewatte, A.S.; Wansapura, J.P.
\newblock Texture analysis of ultrasound images of chronic kidney disease.
\newblock  2017 Seventeenth International Conference on Advances in ICT for
  Emerging Regions (ICTer). IEEE,  2017, pp. 1--5.

\bibitem[Xu \em{et~al.}(2019)Xu, Chang, Su, and Phu]{xu2019classification}
Xu, S.S.D.; Chang, C.C.; Su, C.T.; Phu, P.Q.
\newblock Classification of liver diseases based on ultrasound image texture
  features.
\newblock {\em Applied Sciences} {\bf 2019}, {\em 9},~342.

\bibitem[Sharma and Kumar(2016)]{sharma2016survey}
Sharma, H.; Kumar, S.
\newblock A survey on decision tree algorithms of classification in data
  mining.
\newblock {\em International Journal of Science and Research (IJSR)} {\bf
  2016}, {\em 5},~2094--2097.

\bibitem[Myles \em{et~al.}(2004)Myles, Feudale, Liu, Woody, and
  Brown]{myles2004introduction}
Myles, A.J.; Feudale, R.N.; Liu, Y.; Woody, N.A.; Brown, S.D.
\newblock An introduction to decision tree modeling.
\newblock {\em Journal of Chemometrics: A Journal of the Chemometrics Society}
  {\bf 2004}, {\em 18},~275--285.

\bibitem[Safavian and Landgrebe(1991)]{safavian1991survey}
Safavian, S.R.; Landgrebe, D.
\newblock A survey of decision tree classifier methodology.
\newblock {\em IEEE transactions on systems, man, and cybernetics} {\bf 1991},
  {\em 21},~660--674.

\bibitem[Ke \em{et~al.}(2017)Ke, Meng, Finley, Wang, Chen, Ma, Ye, and
  Liu]{ke2017lightgbm}
Ke, G.; Meng, Q.; Finley, T.; Wang, T.; Chen, W.; Ma, W.; Ye, Q.; Liu, T.Y.
\newblock Lightgbm: A highly efficient gradient boosting decision tree.
\newblock {\em Advances in neural information processing systems} {\bf 2017},
  {\em 30},~3146--3154.

\bibitem[Rezazadeh(2020)]{rezazadeh2020generalized}
Rezazadeh, A.
\newblock A Generalized Flow for B2B Sales Predictive Modeling: An Azure
  Machine-Learning Approach.
\newblock {\em Forecasting} {\bf 2020}, {\em 2},~267--283.

\bibitem[Chen \em{et~al.}(2019)Chen, Zhang, Ma, and Yu]{chen2019lightgbm}
Chen, C.; Zhang, Q.; Ma, Q.; Yu, B.
\newblock LightGBM-PPI: Predicting protein-protein interactions through
  LightGBM with multi-information fusion.
\newblock {\em chemometrics and intelligent laboratory systems} {\bf 2019},
  {\em 191},~54--64.

\bibitem[Sun \em{et~al.}(2020)Sun, Liu, and Sima]{sun2020novel}
Sun, X.; Liu, M.; Sima, Z.
\newblock A novel cryptocurrency price trend forecasting model based on
  LightGBM.
\newblock {\em Finance Research Letters} {\bf 2020}, {\em 32},~101084.

\bibitem[Hossin and Sulaiman(2015)]{hossin2015review}
Hossin, M.; Sulaiman, M.N.
\newblock A review on evaluation metrics for data classification evaluations.
\newblock {\em International journal of data mining \& knowledge management
  process} {\bf 2015}, {\em 5},~1.

\bibitem[Lundberg \em{et~al.}(2018)Lundberg, Erion, and
  Lee]{lundberg2018consistent}
Lundberg, S.M.; Erion, G.G.; Lee, S.I.
\newblock Consistent individualized feature attribution for tree ensembles.
\newblock {\em arXiv preprint arXiv:1802.03888} {\bf 2018}.

\end{thebibliography}

\end{document}